# Beam Diagnostic Examples from High Energy Colliders

*Rhodri Jones*
CERN, Geneva, Switzerland

**Abstract**
This chapter takes a look at how beam diagnostic systems can be used to commission, optimise and solve issues on high energy colliders.

## 1    Introduction

Beam instrumentation and diagnostic systems play an important part in the commissioning and reliable operation of all high energy colliders. As most of these machines are "one of a kind", each brings with it its own, new challenges, for which a comprehensive suite of beam instrumentation is essential to understand issues and optimise performance. This contribution will look at examples of such challenges and how beam instrumentation has helped to overcome them.

## 2    Challenges for the Operation of the LHC

Three of the main challenges for operating the LHC, where beam diagnostics can have an impact are:

*Operating at high energy.* After several years of running at a beam energy of 3.5 TeV and 4 TeV the LHC started running at a top energy of 6.5 TeV in 2015, close to its nominal beam energy of 7 TeV. At this energy the quench thresholds for the superconducting magnets (determining the temperature at which the superconducting coils become normal conducting) are significantly reduced, meaning that the machine will tolerate much less beam loss. As can be seen from Fig. 1, a single pilot bunch of $5 \times 10^9$ protons is already close to the material damage threshold at this energy, while a quench can be induced if less than one millionth of the nominal beam is lost over a 10 ms period.

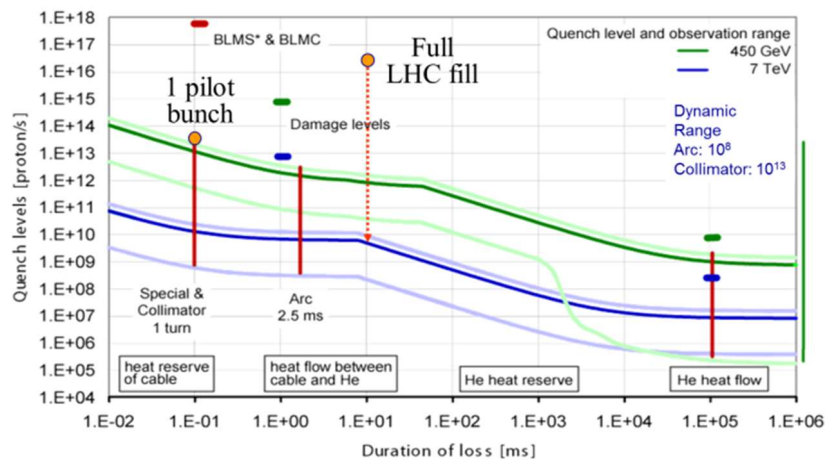

**Fig. 1:** Estimated quench levels for the LHC at injection and top energy.

- **_Running with 25 ns bunch spacing._** During its early years the LHC was operating with a nominal bunch spacing of 50 ns. Such a beam does not suffer from electron cloud build-up and subsequent instabilities, allowing high intensity bunches to be accelerated and collided with little loss. However, such a high intensity per bunch meant that the LHC experiments had to disentangle on average over 30 physics events per bunch crossing. In going to higher energy and tighter focussing, this "pile-up" would become unacceptably large and hence operation at 25 ns spacing (as originally foreseen) was required. This brings with it the problems related to enhanced electron cloud formation and beam induced RF heating of accelerator components.
- **_Coping with high brightness beams._** The combination of high bunch intensity and high energy leads to very high brightness (transversely small, high charge) beams. This poses severe issues for the measurement of their emittances, required for the understanding and optimisation of the machine.

Each of these challenges will now be looked at in detail, with the role played by beam instrumentation and diagnostics in addressing them highlighted.

## 2.1 Operating at higher energy

The main challenge of operating the LHC at high energy is dealing with the reduced quench thresholds of the main dipole and quadrupole magnets. This poses two questions related to managing beam losses: how to deal with any unidentified falling objects interacting with the intense proton beam and how to ensure efficient collimation?

### 2.1.1 Dealing with Unidentified Falling Objects

Initial LHC running was plagued by Unidentified Falling Objects (UFOs) creating beam losses that were large enough to trigger a beam abort by the Beam Loss Monitor (BLM) system [1, 2]. In 2012, 20 beam dumps were identified to be associated with UFOs, with 14 of these occurring at 4TeV. In addition, some 17,000 candidate events were attributed to UFOs below the BLM threshold. The origin of these events is unclear, but as they appear at nearly all locations around the LHC they are believed to be the result of dust particles dropping into the vacuum chamber. The reduction of quench thresholds with energy means that when ramping to 6.5 – 7 TeV many more beam aborts were expected. To buy some additional margin a relocation of the ionisation chambers of the beam loss monitoring system was carried out during the first LHC Long Shutdown, LS1 2013-2015 (Fig. 2).

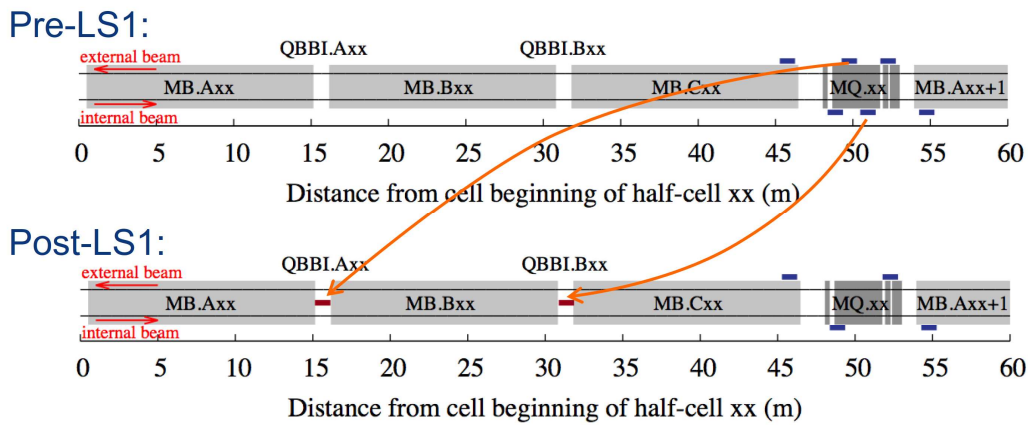

**Fig. 2:** Relocation of LHC beam loss system ionisation chambers (blue) to optimise protection against UFOs. MQ - main quadrupole: MB - main bend (dipole).

Designed to protect the superconducting magnets from beam losses at maximum beta-function locations, the original system had 3 ionisation chambers per beam and per quadrupole. However, as UFOs can occur at any location, an event in one of the dipoles between two quadrupoles will result in a relatively small signal by the time the secondary shower reaches the beam loss monitor sitting on the quadrupole. In order to avoid the dipole quenching with this configuration, the BLM thresholds have to be substantially reduced, well below what is required to avoid quenches due to 'standard' beam loss at the quadrupoles. After significant simulation work it was decided to move two of the BLMs from the quadrupole to the dipoles. This was found to leave sufficient redundancy to avoid quenches due to losses at the quadrupoles while gaining a factor 30 in sensitivity to UFO events occurring in the dipoles.

The relocation of these BLMs thus enabled the thresholds be set such as to abort the beam and avoid quenching the magnets for beam losses and large UFO events at 6.5-7 TeV, while limiting unnecessary beam aborts from small UFOs that would not result in a magnet quench.

### 2.1.2    Ensuring Efficient Collimation

The beam cleaning efficiency necessary to ensure that any beam loss in the cold LHC arc is maintained well below the quench threshold is ensured by a comprehensive collimation system (Fig. 3) [3]. A series of primary, carbon-jawed collimators scatter the transverse and longitudinal beam halo, which is then absorbed by secondary collimators with slightly larger opening. Tungsten-jawed tertiary collimators near the experiments and various protection devices complete the collimator hierarchy, cleaning-up the remainder of the secondary halo and protecting against equipment failure, such as the misfire of the dump kickers. The gap between opposite jaws in the 1.2 m long primary collimators are typically below 2 mm.

Guaranteeing the correct set-up of these ~100 moveable devices and ensuring that the beam remains centred during long periods of operation is the job of the beam loss and beam position systems.

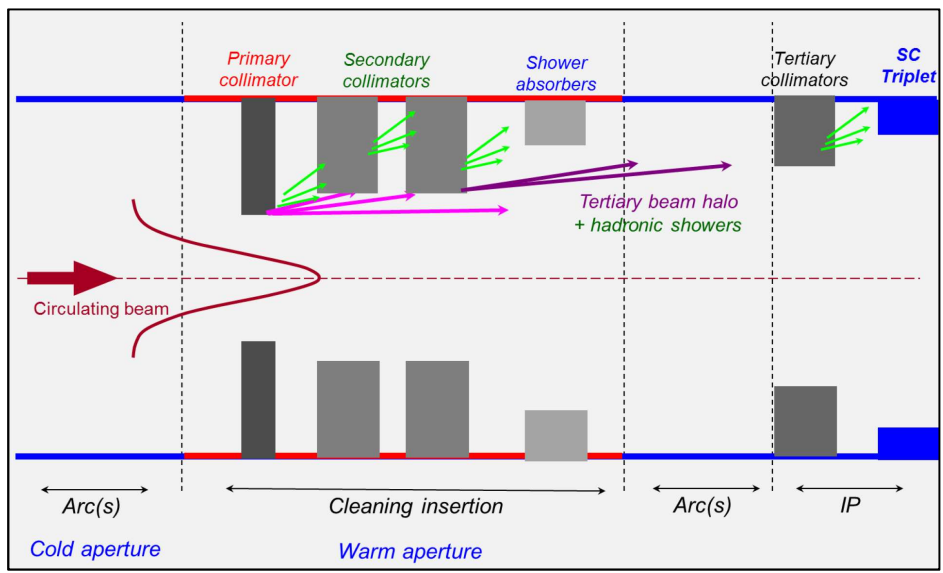

**Fig. 3:** The LHC collimation system.

### 2.1.2.1    *Defining the right collimator positions*

Once a stable orbit has been defined, the setting up of the collimators is carried out using the LHC beam loss system [1], consisting of over 3000 ionisation chambers. With very low intensity beam in the machine, each collimator jaw is individually moved in until it touches the beam halo, inducing a loss spike which is detected in a neighbouring BLM. Initially this procedure took over a day to perform for all collimators, but has since been reduced to less than one hour, thanks to the parallelisation of the task and an increased BLM data stream.

Once centred, the collimation hierarchy is validated by so-called loss maps. This involves creating an artificial loss and monitoring the leakage of lost particles into the cold magnets using the complete BLM system. An example of such a loss map is shown in Fig. 4. It can be seen that the losses are mostly localised in the betatron and momentum cleaning collimation regions. The maximum leakage of lost particles into a cold magnet is less than 0.02 %, within the design specifications.

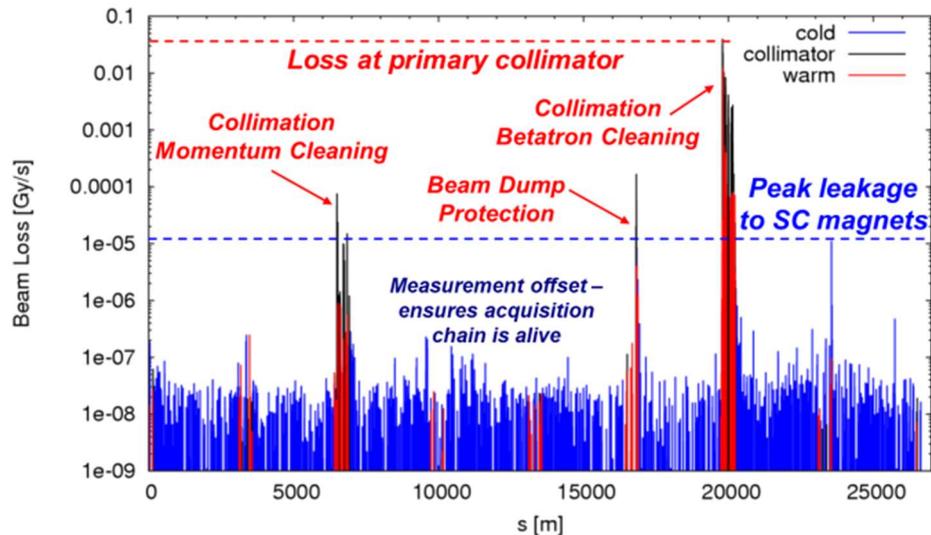

**Fig. 4:** Example of an LHC loss map showing the response from all beam loss monitors around the LHC.

All the original 16 tertiary collimators have now been replaced by a design which includes embedded beam position monitors (BPMs) (Fig. 5) [4]. The beam retraction settings of these collimators are linked to the amount of focussing applied in the experimental insertions. A better knowledge of the beam position at their location allow safety margins to be reduced to allow tighter focussing, and hence higher luminosity.

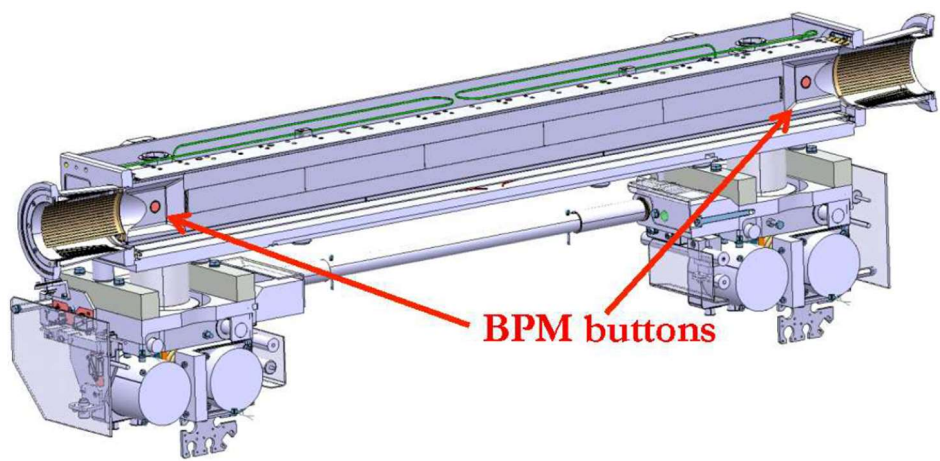

**Fig. 5:** LHC tertiary collimator with beam position monitors embedded within the jaw (only one jaw shown).

Such collimators with embedded BPMs have demonstrated that the time required to centre the collimator around the beam to within 10 μm can be reduced to a mere 20 seconds [5]. The system uses two pairs of button pick-ups, one at the upstream and one at the downstream end of the collimator, slightly retracted from the face of the jaw. The signals are processed using compensated diode peak detector electronics, specially designed to give excellent resolution, < 100 nm, for centred beams [6]. This electronics for accurate, high resolution orbit measurements has been so successful that it has also been extended to work in parallel to the existing bunch-by-bunch orbit and trajectory system in critical locations.

### 2.1.2.2 Maintaining the right collimator positions

Once the right collimator positions have been defined, the role of maintaining the beam in their centres is the job of the LHC beam position system [7]. This is comprised of 1054 beam position monitors, the majority of which (912) are 24 mm button electrode BPMs located in all arc quadrupole cryostats. The remaining BPMs are enlarged (34 mm or 40 mm) button electrode BPMs mainly for the stand alone quadrupoles, or stripline electrode BPMs used either for their directivity in the common beam pipe regions or for their higher signal level in the large diameter vacuum chambers around the dump lines.

The beam position acquisition electronics is split into two parts, an auto-triggered, analogue, position to time normaliser which sits in the tunnel and an integrator/digitiser/processor VME module located on the surface. The link between the two is made using a fibre-optic connection of up to 3 km in length. Each BPM measures in both horizontal and vertical planes, making a total of 2156 channels.

The data from all these channels is fed at 25 Hz to a central orbit feedback system which, using a regularised SVD approach and a closed loop bandwidth of 0.1 Hz, can maintain rms orbit stabilities of better than 70 μm globally and 20 μm in the arcs. The measured fill-to-fill reproducibility when going into collision at the ATLAS interaction point throughout 2012 is shown in Fig. 6. Whilst a slow drift of 80 μm is seen to build-up over the year, the reproducibility from one fill to the next is excellent with the difference only 7 μm rms.

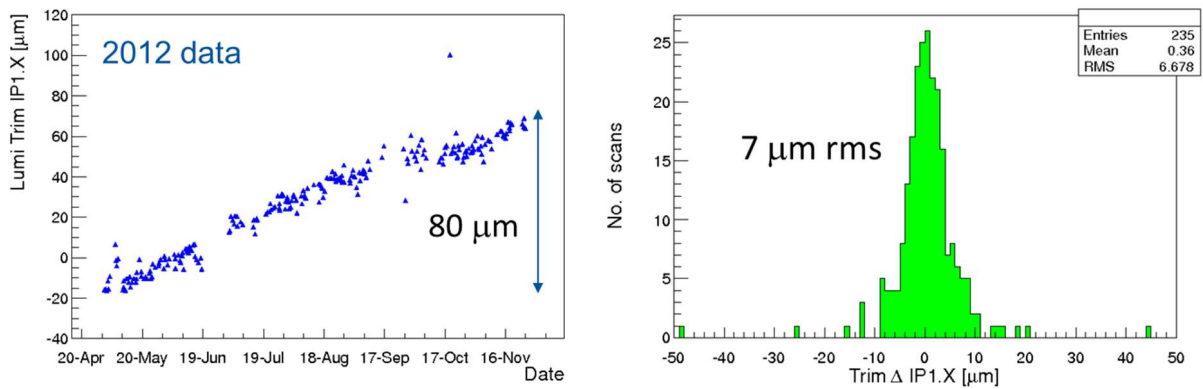

**Fig. 6:** Orbit reproducibility going into collision at the ATLAS experiment as measured in 2012.

The main performance limitation of the orbit feedback is linked to an observed systematic position dependence on temperature that initially caused errors of greater than 300 μm on the orbit measurements. This was reduced to the order of 100 μm by regular calibration and applying temperature corrections to the data. To try and minimise this problem temperature controlled racks were added to house the BPM surface electronics. These maintain a stable temperature to within ±0.2 °C which typically keep any orbit drifts well below 50 μm.

## 2.2 Running with 25ns bunch spacing

The main challenge of running with 25 ns spaced bunches in the LHC is dealing with the electron cloud that this generates [8]. These electrons are created due to secondary emission from the beam-pipe wall, initially via ion bombardment or synchrotron radiation. As these electrons drift into the chamber they are accelerated by a passing bunch, hitting the opposite wall, creating more secondary electrons. This process is repeated with the following bunches, creating and avalanche of electrons, which eventually forms an electron 'cloud'. Apart from inducing a dynamic pressure rise and an additional heat-load for the cryogenic circuits, it also has an impact on beam quality, with the resulting instabilities able to lead to particle loss and emittance growth.

For a given bunch spacing, the threshold at which this cloud can develop is given by the secondary electron yield of the wall material. This can be lowered by scrubbing the surface through the impact of a high density of electrons, which is why many machines introduce scrubbing runs with high intensity beams. Moving from 50 ns spacing to 25 ns spacing significantly lowers the secondary electron yield threshold at which the electron cloud forms. It is therefore important to be capable of observing instabilities, measuring beam loss and monitoring emittance growth to understand, combat and optimise operation with such beams.

### 2.2.1 Bunch-by-Bunch Diagnostics

Understanding the impact of the electron cloud and other instabilities can only be effectively done using bunch-by-bunch measurements. In the LHC nearly all instrumentation systems have therefore been designed to deliver such data.

A new type of wall current transformer was developed [9] to improve the bunch-by-bunch resolution and remove any dependency on beam position and bunch length. This provides a bandwidth > 100 MHz, less than 0.1 %/mm position dependence and less than 0.1 % bunch length dependence. Such monitors are routinely used to measure bunch-by-bunch intensity losses throughout the accelerator cycle, and are an excellent tool to understand and disentangle losses coming from various origins. Fig 7. shows the typical pattern expected from losses induced by electron cloud, with a rise in loss along each batch which becomes more and more marked throughout the fill.

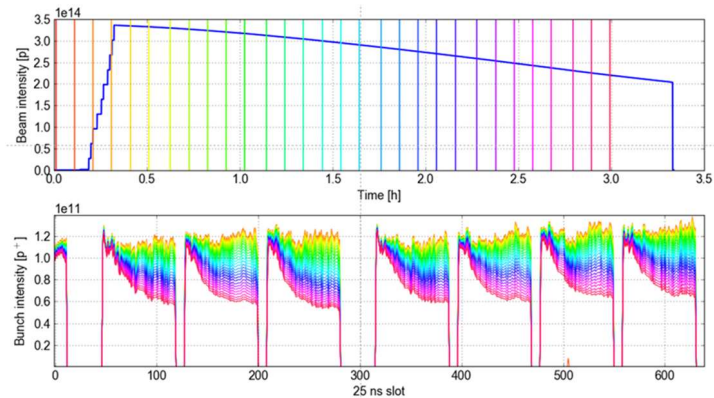

**Fig. 7:** Upper plot: total intensity evolution during a physics fill with colour coded snapshots corresponding to the time of the measurements in the lower plot. Lower plot: Evolution of bunch-by-bunch intensity during a physics fill as measured by a wall current transformer. Electron cloud creates instabilities along the bunch trains leading to this characteristic loss pattern.

A synchrotron light monitor is also capable of providing bunch-by-bunch diagnostics and is extensively used in the LHC to study the effect of electron cloud and other phenomena acting on the emittance of individual bunches. To achieve such measurements an image intensifier, typically based on a multichannel plate, is inserted between the photocathode and the camera sensor. This intensifier is capable of being switched on and off on a nanosecond timescale, allowing the camera to be gated on a single bunch. By acquiring successive images on different bunches a complete picture of the transverse size of each bunch can be built-up. With a readout rate of 50 frames/s, capturing all of the LHC's nearly 3000 bunches only takes a minute.

Fig. 8 shows an example of bunch-by-bunch emittance measurements when conditioning the machine against electron cloud effects after a long maintenance shutdown. Instabilities leading to increased emittance are clearly visible in both planes towards the end of the injected batches. These measurements are invaluable in quantifying improvements made during the course of this conditioning, and are also used to detect and correct non-uniformity in the emittance of the beam coming from the LHC injectors.

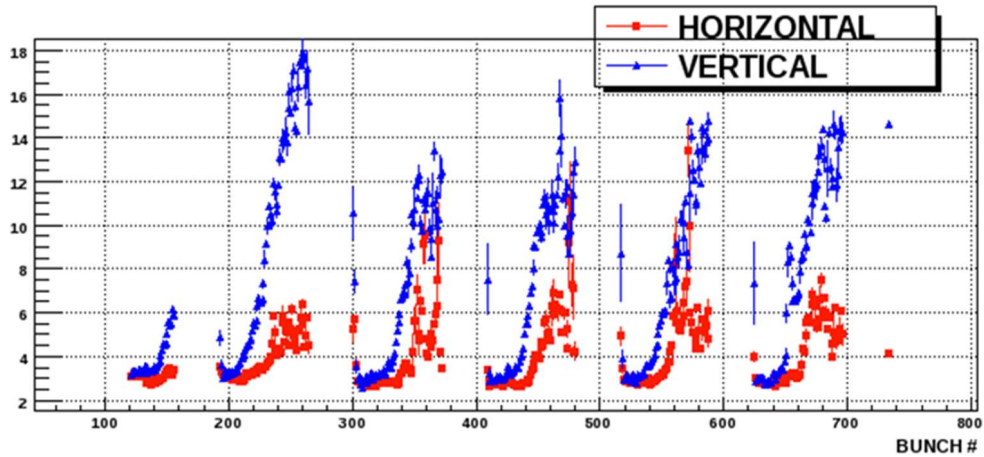

**Fig. 8:** Bunch-by-bunch emittance, as calculated from the beam profiles measured by the synchrotron light monitor, showing the blow-up of bunches towards the tail of each injected batch due to electron cloud instabilities.

### 2.2.2 Intra-Bunch Diagnostics

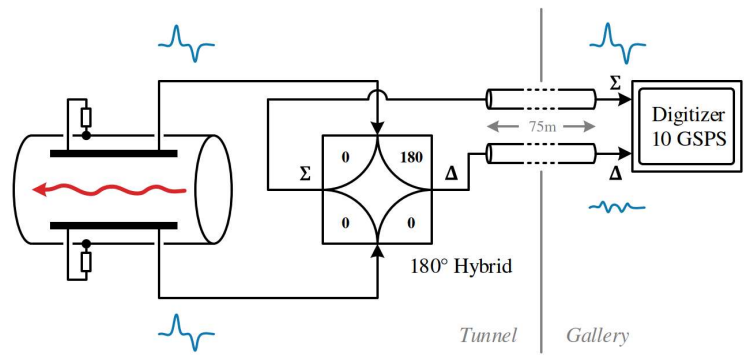

**Fig. 9:** The principle of the LHC head-tail monitor. The sum ($\Sigma$) and difference ($\Delta$) of the electrode signals of a long "stripline" type BPM are obtained using a passive 180° hybrid and sampled using a high speed digitizer located in a service gallery.

Another important diagnostic to understand the origin of instabilities is intra-bunch measurement. Various instabilities have been observed in the LHC during the beta squeeze (when focussing the beam in the experiments) and whilst going into collision, leading to beam loss and emittance growth. Disentangling the many possible causes of these instabilities requires a detailed knowledge of their effect on the beam. The main instrument for characterizing intra-bunch beam instabilities in the LHC is the so-called "head-tail monitor" [10]. The system, shown as a simplified block diagram in Fig. 9, is based on the high speed acquisition of a long "stripline" type beam position monitor. The length of the stripline is chosen so as to separate the 'signal' and its 'reflection', using the fact that consecutive bunches are spaced in time by much more than the bunch length. Since the signal and its reflected pulse are well separated in time the reflected pulse can be removed by gating the signal after digitisation in the time domain, effectively leaving just the intra-bunch position variation along the bunch.

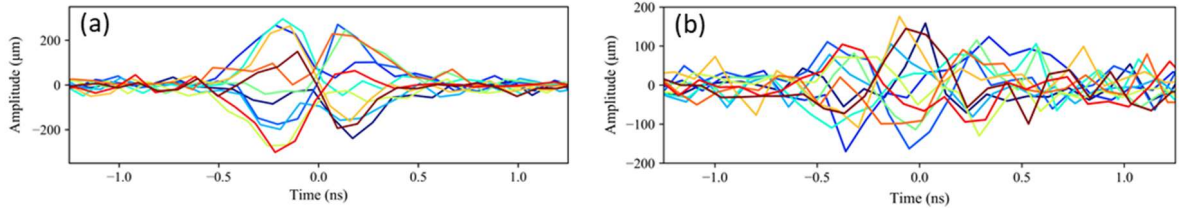

**Fig. 10:** Example of a Mode 1 (a) and Mode 4 (b) instability captured by the LHC "head-tail" monitor.

The LHC head-tail monitor has two dedicated, single-plane, 40 cm long stripline beam position monitors per beam, one horizontal and one vertical, installed at locations with high β-functions in the measurement plane. A commercial, wideband, 180° hybrid produces the analogue sum (Σ) and difference (Δ) of each pair of electrode signals. The signals are directly digitized with a high-speed digitizer located close to the beam line in a service gallery. Using these systems intra-bunch instabilities up to mode 4 have been clearly distinguished (see Fig. 10). These signatures can be used by beam physicists to determine the origin of the instability through comparison to simulation or by dedicated machine development experiments.

## 2.3 Coping with high brightness beams - Measuring Small Beam Sizes

The LHC is equipped with 4 beam size measurement devices: optical transition radiation screens for the setting-up of injection and extraction; wire-scanners for absolute measurement and the calibration of the other devices; synchrotron light monitors; rest gas ionisation monitors. Each of the three devices used with circulating beam currently present limitations when measuring high brightness beams.

### 2.3.1 Wire-Scanners

The operational limits for the LHC wire-scanners are defined by their wire-breakage limit due to beam heating given, for the carbon wires used in the LHC, by the process of wire sublimation (Fig. 11) [11]. At the injection energy of 450GeV the limit sits at a total intensity of around $2.7\times10^{13}$ protons. This is not even sufficient to measure a full injected SPS batch of 288 bunches with 25ns spacing. At 6.5TeV, the calculated limit of $2.7\times10^{12}$ protons is a mere 20 bunches. Dedicated runs are therefore required with a low number of bunches when using the wire-scanners to cross calibrate the other devices.

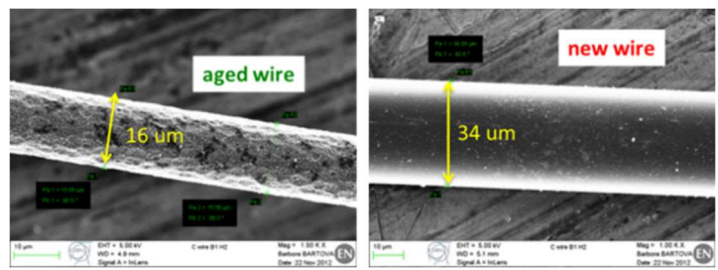

**Fig. 11:** The effect of sublimation on the carbon wire of the LHC wire-scanners.

### 2.3.2 Synchrotron Light Monitor

At top the energy of 6.5-7 TeV the imaging resolution from the synchrotron light monitor is dominated by diffraction. The optical system is therefore designed to use ultra-violet (UV) compatible optics and a CCD camera with a UV sensitive photocathode [12]. Even so, when imaging in a narrow band at 250 nm, the contribution from diffraction is estimated to be ~250 µm compared to a beam size of only 180 µm. A good understanding of diffraction effects and all other distortions is therefore necessary to extract an accurate absolute beam size from these images. As this is difficult to obtain, the system has to rely on cross calibration with the wirescanners at both injection and top energy.

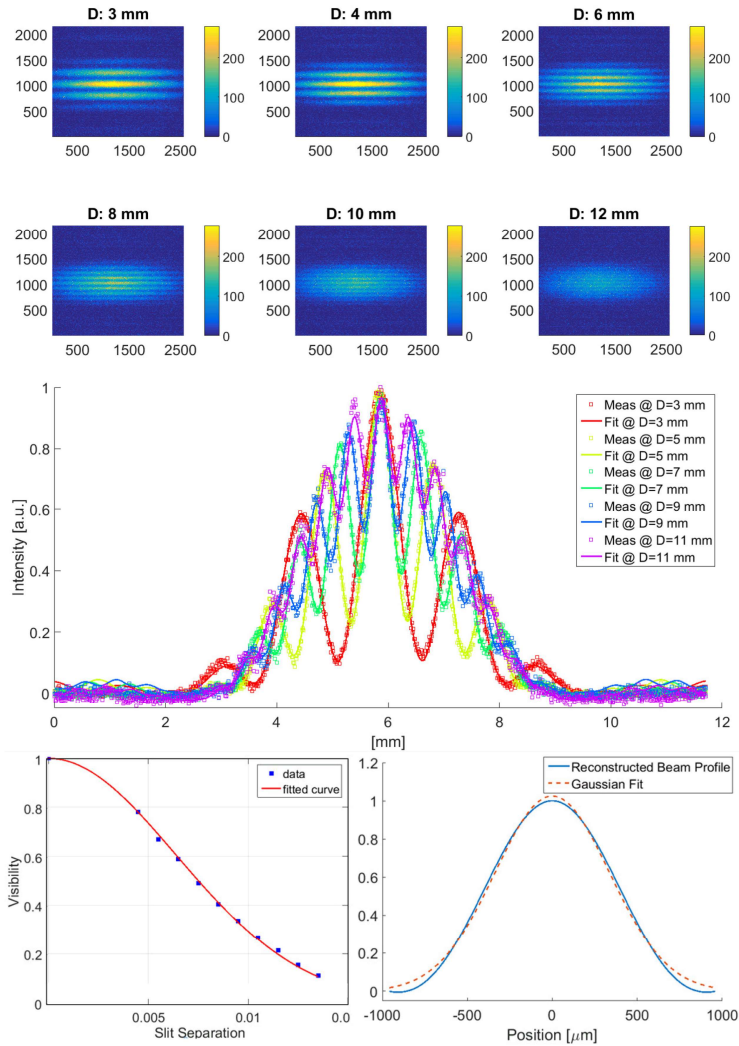

**Fig. 12:** Top: interferograms recorded for various slit separation using a 560 nm filter. Middle: projections allowing the visibility to be calculated. Bottom: plot of visibility against slit separation from which an inverse Fourier transform leads to the reconstructed beam profile.

Due to the extent to which imaging is dominated by diffraction at 6.5-7TeV a new optical line has been added to perform interferometry, though a collaboration with KEK (Japan), SLAC (US) and CELLS-ALBA (Spain) [13]. This non-diffraction limited technique is widely used in electron machines with very small beam sizes and relies on the fact that the visibility of the interference pattern is dependent on the beam size. The results of LHC measurements at top energy are shown in Fig. 12. A scaling factor of about 1.3 is yet to be understood between interferometry and imaging at top energy, suspected to come from the complicated LHC source geometry. Nevertheless interferometry was able to provide coherent relative bunch size measurements.

### 2.3.3    Ionisation Profile Monitor

The LHC ionisation profile monitor (IPM) is based on electron collection using a 0.2 T guide magnet, a multi-channel plate for amplification and an optical readout from a phosphor screen with a radiation-hard camera [14]. While this monitor has worked well with $Pb^{54+}$ ion beams (its primary purpose being to measure such beams as they emit very little synchrotron light at the LHC injection energy of 450GeV), it was seen to suffer from severe image distortion during the proton energy ramp. This was suspected to be due to space-charge effects from the high brightness proton beams, something that was confirmed by simulations. The distorted profile (Fig. 13) cannot currently be deconvoluted to extract the original profile, and the only real solution to this problem seems to be to increase the magnetic guide field to around 0.7 T. In addition these monitors were seen to suffer from beam induced RF heating and had to be removed during the first long LHC shutdown.

Plans are in the pipeline to replace these detectors with a version based on hybrid pixel technology, as recently successfully implemented in the CERN PS machine [15]. Such technology overcomes many of the disadvantages of the original microchannel plate monitors, such as ageing effects and the need for gas injection, while allowing bunch by bunch diagnostics. These monitors would, nevertheless, still need to be coupled to a stronger magnetic guide field to overcome space charge effects with high brightness proton beams.

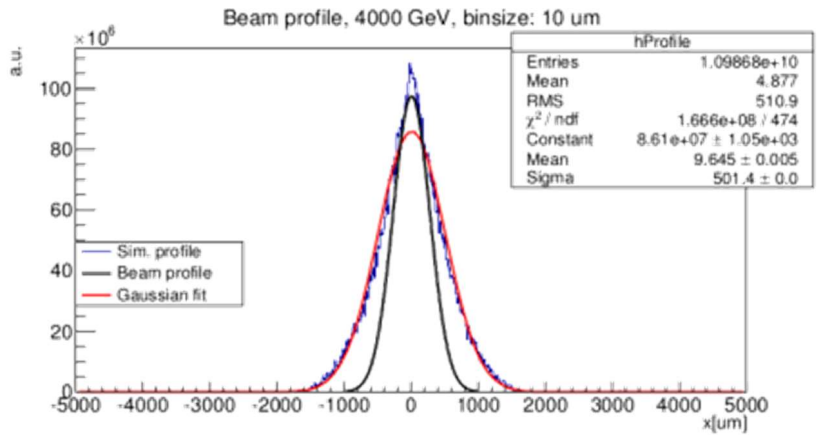

**Fig. 13:** Profile as measured by the IPM at 4TeV showing the large distortion due to space charge.

## 3    Measuring the Machine Optics Functions

The measurement and correction of optics parameters has been an area of intensive study since the advent of strong focusing synchrotron accelerators, where perturbations from field imperfections and misalignments became a concern. Traditionally, colliders have led the development of methods for optics control based on turn-by-turn centroid position data, while lepton storage rings have focused on closed-orbit-response techniques [16]. Both of these methods rely heavily on the use of the beam position system of the accelerator, and are now often driving its requirements.

## 3.1 Optics Measurement using Turn-by-turn Techniques

In 1983 a major achievement took place in the CERN-ISR, where the Beam Position Monitors (BPMs) were used successively around the collider to measure the relative amplitude and phase advance of the β-function by observing the amplitude and phase of induced betatron oscillations [17]. This was the first time that machine optics had successfully been reconstructed from individual BPM data, at that time using a BPM system that was entirely based on analogue electronic technology.

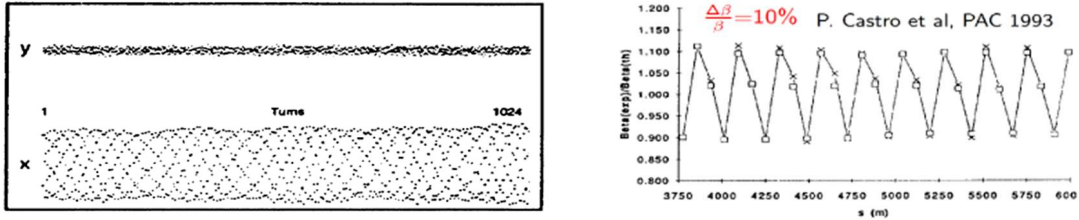

**Fig. 14:** LEP β-beating example. Left: turn-by-turn BPM data. Right: Δβ/β for a section of the machine.

The first optics measurements using digital, turn-by-turn BPM data were performed at CERN-LEP (Fig. 14) [18]. The β-function at each BPM location was extracted from the phase advance between 3 BPMs, assuming a good knowledge of the focusing elements in between:

$$\beta_{measured}^{BPM1} = \beta_{model}^{BPM1} \left( \frac{\{cot\varphi_{12} - cot\varphi_{13}\}_{measured}}{\{cot\varphi_{12} - cot\varphi_{13}\}_{model}} \right)$$

This method, known as "β from phase", was also used in CESR (Cornell, USA) in 2000 to minimize the β-beating, the difference between the measured β and the design β (Δβ), to reach an rms difference of only 2% [19]. This is still one of the best optics correction achieved in a lepton collider.

A limitation of this method is its reliance on good quality BPM data. Identifying BPMs giving poor readings or BPMs with excessive noise is therefore very important. A major step forward in achieving a more robust analysis was taken at SLAC in 1999, where singular value decomposition (SVD) techniques were used to isolate faulty BPMs and identify noise components affecting the oscillation data [20]. To further mitigate this issue the 3 BPM method has recently been extended to take into account any number of BPMs [21], resulting in a much better overall resolution in the measurement of the β-function.

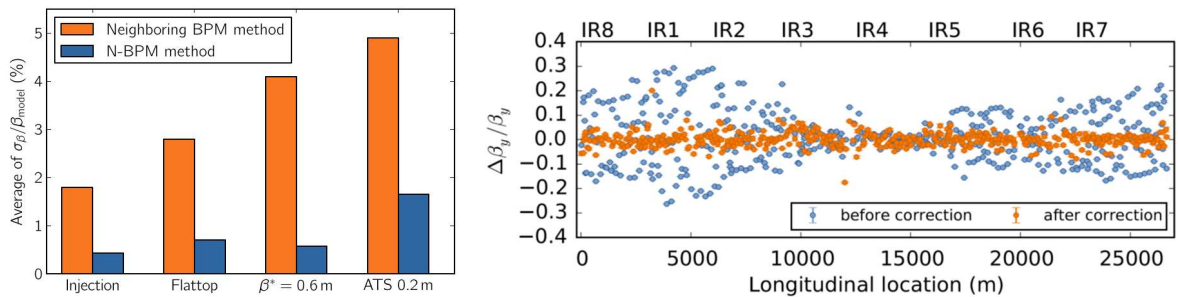

**Fig. 25:** Left: Comparison for the average error bar on the experimentally measured β-function for different optics configurations at 4 TeV between the "old" 3 neighbouring BPM and "new" N-BPM methods. Right: Optics correction example from the LHC in 2015.

In order to initiate a sufficiently large centroid motion to be visible on a turn-by-turn basis with a BPM system, the beams typically need to be kicked to relatively high amplitude using a fast, electro-magnetic kicker. The disadvantage of this technique is that in a non-linear machine these oscillations are quickly damped. This damping is a big drawback for optics measurement as the amount of useful BPM data available is then highly limited. In addition, for hadron machines, this leads to significant emittance blow-up. An alternative excitation technique uses an "AC Dipole" excitation, originally developed at RHIC (BNL, USA) for crossing polarisation resonances [22]. Here, a forced oscillation is put onto the beam near the betatron tune, but outside the frequency range covered by the tune spread. If performed adiabatically, this leads to a steady, high amplitude oscillation without emittance blow-up, which is excellent for turn-by-turn optics measurements (see Fig. 16).

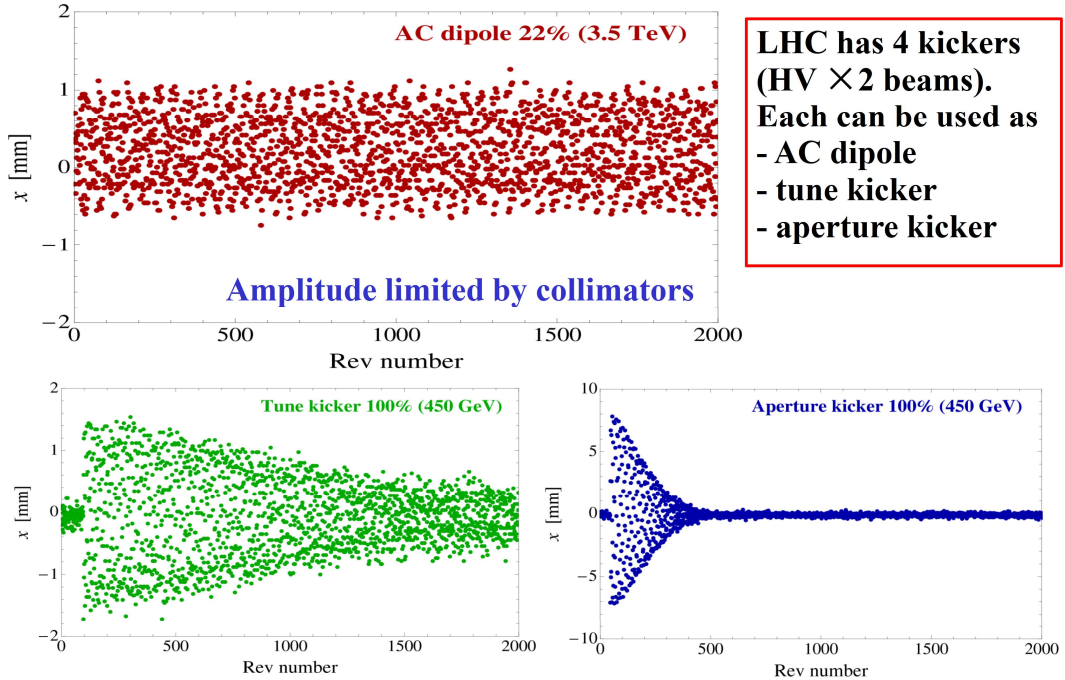

**Fig. 16:** Examples of excitation for optics measurements in the LHC. Top: AC-dipole excitation. Bottom Left: Single kick at injection energy using the tune kicker. Bottom Right: Single kick at injection energy using the aperture kicker.

### 3.2 Beam Instrumentation Challenges for Improved Optics Measurements

There are three main challenges to improving our understanding of the machine optics functions using turn-by-turn techniques:

- The excitation level needs to be reduced in order to allow operation under tight machine protection constraints (also limiting any emittance growth in hadron machines), or to avoid non-linearities due to strong sextupoles in next generation synchrotron light sources. This implies improving the turn-by-turn resolution of the BPM systems in order to obtain the same signal to noise performance for smaller excitation levels.

- Secondly, optics measurements would benefit from a much better BPM linearity in the range of the excitation, and in the overall calibration from BPM to BPM. Light sources are currently at the 1-2% level, the LHC at the 3-4% level. Improving this to below the 1% level would allow the use of the oscillation amplitude as well as the phase information for β-function reconstruction.

- Thirdly, all machines would benefit from a better BPM design to lower the coupling impedance these BPMs present to the beam. This is a serious issue for synchrotron light sources where the machines becomes more sensitive to collective effects as lower beam emittances are achieved, with the BPMs accounting for a significant fraction of the total impedance budget. In addition, the short range, high frequency wakes induced by the BPMs can result in beam induced heating.

## 4    Beam Dynamics Studies using Betatron Tune Spectra

Betatron tune measurements are useful for a variety of accelerator physics applications. The tune shift with quadrupole strength gives the local beta function, the tune shift with RF modulation the chromaticity, the tune shift with beam current the transverse impedance and the tune shift with amplitude the strength of non-linear fields. Comprehending these tune spectra is also important for the optimisation of beam lifetime, limiting emittance growth, and reducing beam losses through the understanding of instabilities, space charge effects, beam-beam interactions etc.

A typical tune spectrum has several constituents. As the measurement is usually taken using a single BPM in the ring, the main components are revolution lines generated by the periodicity of the circulating beam with each revolution line having tune sidebands. The revolution lines are usually filtered out by the front-end electronics or simply not displayed in the spectrum reported in the control room, leaving just the tune from coherent betatron motion in the plane of excitation. This is typically displayed in fractional tune units, from 0 to 0.5 (or 0.5 to 1) of the revolution frequency. If coupling between the transverse planes is present a second peak at the tune frequency corresponding to the non-excited plane will also be visible. An example of the beam response to single kick excitation and the corresponding frequency spectrum is shown in Fig. 17.

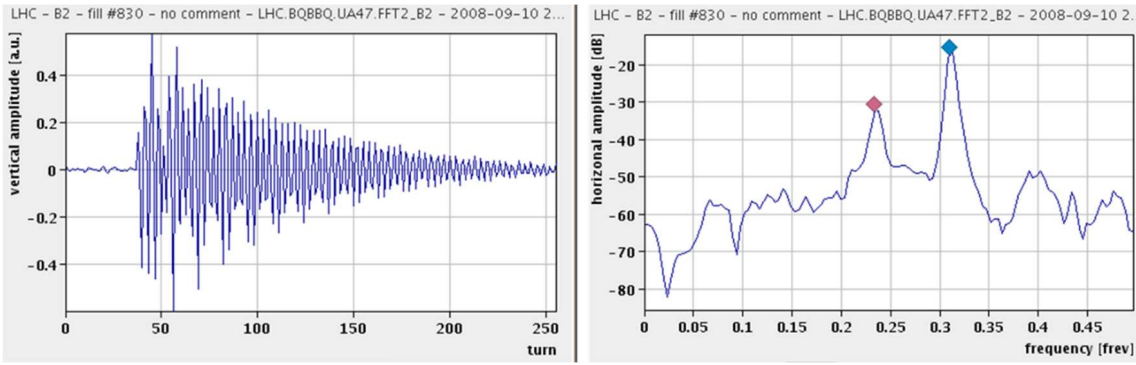

**Fig. 17:** BPM turn-by-turn response to a beam excited with a single kick (left). Corresponding frequency spectrum showing the presence of coupling (right).

In the presence of synchrotron motion the interplay between longitudinal motion and transverse betatron motion of an ensemble of particles leads to an amplitude modulation of the centroid of the particle bunch, as measured by a BPM, which depends on chromaticity. This manifests itself as additional sidebands that appear on either side of the main tune peak in the spectrum. The distance of these sidebands from the tune can be modified by impedance and space charge effects and hence provides important information for optimising machine performance [23]. Fig. 18 shows the horizontal tune spectra obtained with a $U^{73+}$ beam at GSI at three different intensities, along with the predicted head-tail mode shifts as a function of the space charge parameter. At low intensity the synchrotron peaks are almost equidistant, but as the intensity is increased and space charge effects come into play, these synchrotron satellites change both their frequency and amplitude. By studying these spectra and comparing them to simulation it is possible to disentangle beam dynamic effects coming from various sources.

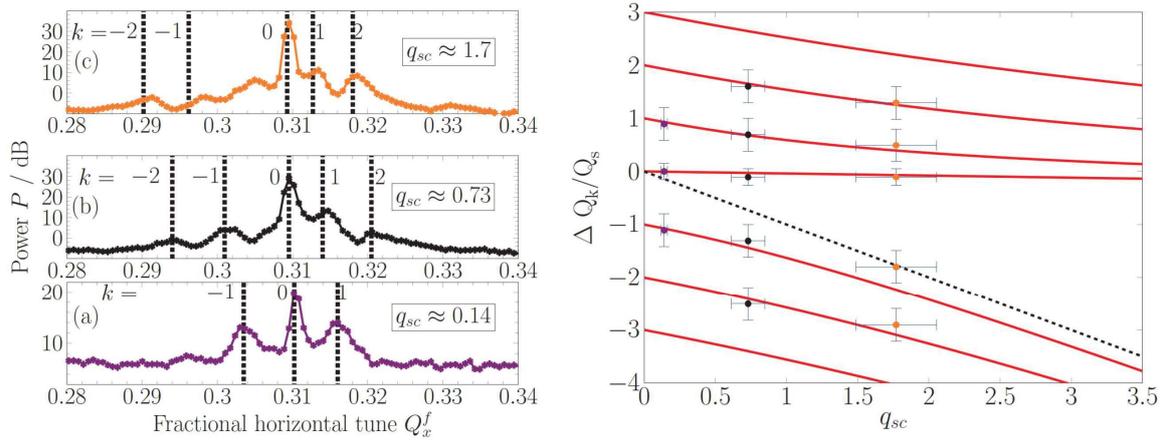

**Fig. 18:** Left: Tune spectra at the GSI SIS18 for various beam intensities, showing the effect of space charge on the synchrotron sidebands. Right: simulated and measured synchrotron sideband separation for various space charge regimes.

# 5 Some Examples of Beam Diagnostics in Action

This section is meant to serve as general entertainment for those readers who have made it this far! Two old examples of beam diagnostics in action have been selected from CERN-LEP operation, and show how difficult it can be to interpret primary measurements and decide on the right actions for solving a problem in an accelerator.

## 5.1 The CERN-LEP beam does not circulate!

The LEP accelerator had a very regular operation schedule. Every year it was used for about 8 months for physics, followed by a 4 month maintenance and upgrade shutdown. During this shutdown major intervention work was sometimes carried out on the machine. At the next start-up it was often expected that typical problems, such as inverted magnet polarities, would have to be overcome. One year the start-up was particularly bad, with neither the electron beam nor the positron beam capable of being made to circulate. Several hours were devoted to checking all vacuum conditions, power supply currents, settings of the radio frequency system, injection deflectors and so on, but nothing indicated a severe problem. Finally people started to look in detail at the measured beam trajectory from the injection point onwards. A typical example for the positron beam is shown in Fig 19.

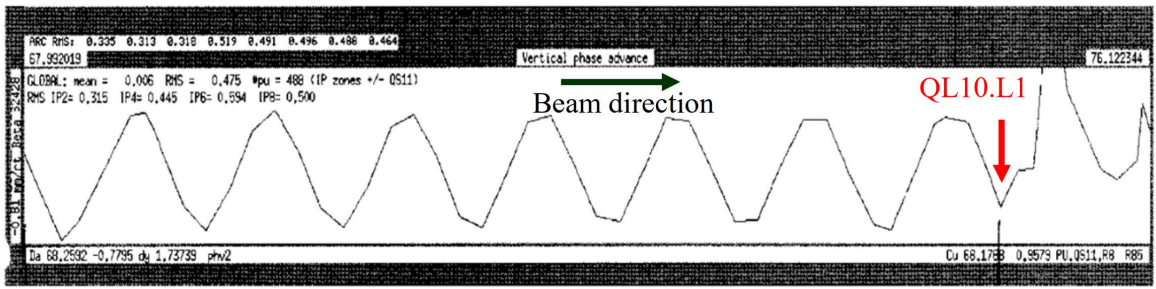

**Fig. 19:** Measurement of the LEP phase advance when beams did not circulate

What is actually shown in Fig. 19 is the phase advance from one beam position monitor to the next, as calculated from the measured beam trajectory. At a particular quadrupole (QL10.L1) the regular pattern is distorted. Additional measurements also indicated that most of the beam was lost at this point. The first conclusion was to suspect a problem with this quadrupole. People went in, measured the current in the quadrupole, checked its polarity, inspected its coils, but could not find anything abnormal.

The indications of the beam measurements, however, clearly pointed to a problem at this location. After many discussions and potential hypotheses it was decided to open the vacuum chamber. It should be noted that this was a major intervention, causing a stop of the accelerator for several days. One can understand the surprise of the intervention team when they looked into the open vacuum chamber and saw a beer bottle!!! During the shutdown intervention, somebody had sabotaged the LEP accelerator and inserted a beer bottle into the beam pipe (Fig. 20)! What had upset the operation team most at the time was the fact that it was a very unsociable form of sabotage - the bottle was empty!

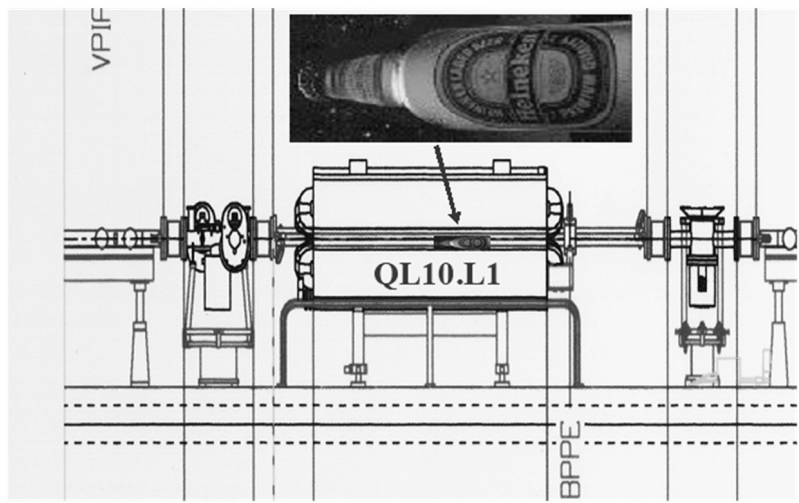

**Fig. 20:** The mystery of the beam circulation problem in LEP is solved!

## 5.2 The CERN-LEP beam gets lost during the beta squeeze

This again is one of the stories from LEP operation which took several hours using beam diagnostics to solve. The problem in itself is pretty complex, and therefore requires some additional explanations beforehand.

The acceleration of the particle beams and the change of the lattice function in the insertion regions in order to get smaller values of the beta-function at the crossing point (hence higher luminosity) are so-called "dynamic processes". The presence of the beam requires that all actions are well synchronised. For example, the power converters of all relevant magnetic circuits have to be controlled such that beam parameters like the closed orbit, tunes and chromaticities stay within tolerance during the dynamic process. In order to achieve this, the behaviour of these beam parameters is periodically measured as a function of time and the corresponding power converter tables are updated.

During one period of LEP operation it was found that the beams were lost during the beta squeeze. Shortly before the total loss of the beams a significant beam loss was measured. As standard practice when encountering such problems, the engineer in charge (EIC) launched a new machine cycle with diagnostic facilities such as "tune history" (the measurement of the betatron tunes as a function of time) switched on. This indicated that the vertical tune moved out of tolerance during the beta squeeze. Fig. 21 shows an excerpt from the actual LEP logbook entry of this event.

As a result of this observation, the EIC launched another cycle, but inserted a breakpoint (to stop the accelerator cycle) just before the critical moment in the beta squeeze when the deviation in tune occurred. Having reached that breakpoint the tunes were measured statically and found to be perfectly within tolerance. The beta squeeze was then executed step by step, and to the big surprise of the operations crew, the tunes were found to be correct at all times. The beam had passed the beta squeeze like on an ordinary day! But on the next attempt, without a break in the cycle, the beam was again lost at the same moment, and several people scratched their heads to find an explanation.

Straight through to 96 GeV.
At ~97-98 GeV    e⁻  lage vertical oscillation
OPAL trigger.  Maybe a bit too ambitious
Tunehistory    01-12-40    fill 7065
→ nothing particularly nasty.

Big radiation spikes in all expts.

01:40    22 GeV    4QSO.    Breakpoint at 93 GeV.
640 µA    .234 /.164        5·27 mA
93 GeV    4QSO            01-58-36    VRMS ~c
Tunehistory    01-50-25    fill 7066

**Fig. 21:** Excerpt from the LEP logbook when beams were lost during the beta squeeze

Finally, the following measurement was made. The machine was prepared and a breakpoint again inserted just before the critical beam loss. Once this point was reached, the EIC requested the execution of one further step in the beta squeeze. The facility by which one could execute a single step in a dynamic process had the additional feature that one could specify the rate of current change of any machine element. This current rate limitation was changed from 5A/s (nominal) down to 2.5 A/s on consecutive steps. The corresponding tune history (the result from the vertical plane is plotted on the lower graph) is shown in Fig 22.

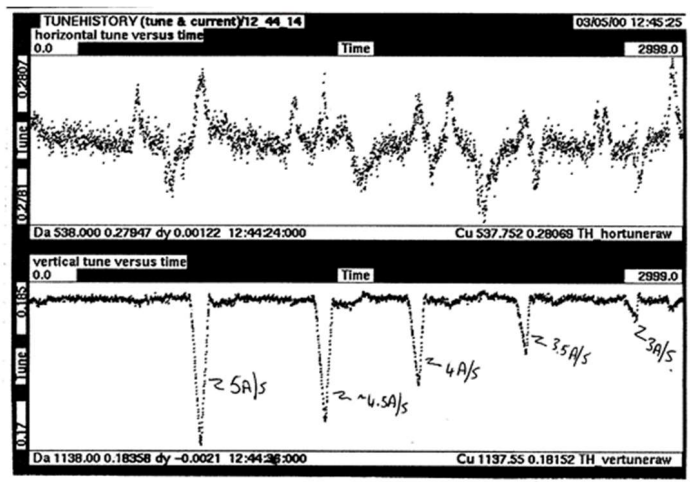

**Fig. 41:** The LEP tune history during the beta squeeze for various power converter ramp rates

One can clearly see that a huge (negative) tune excursion occurred when the step was executed at the nominal rate. This observation led the EIC to the right conclusion, which was that one of the power supplies was able to deliver the demanded current statically, but not dynamically. When this was discussed with experts from the power converter group, they indicated that the power supplies for the superconducting insertion quadrupoles were built as two blocks in series, each of them able to deliver the necessary current (each block typically 1000 A/10 V). Both of these blocks were required to have enough voltage margin to enforce a current change against the inductance of the quadrupole coil. This then explained the whole story. One of these blocks was faulty, but since the remaining working block could deliver its static current, it was not detected by an alarm or surveillance circuit. If the dynamic rate was too high, however, this single block could not provide enough current leading it to lose synchronism with the other power converters. This resulted in the large tune change observed and ultimately the total loss of the beam.

These two examples show the enormous potential of beam instrumentation if they are used in the right combination by intelligent people.